\documentclass[runningheads]{svmult}

\usepackage{makeidx}   
\usepackage{graphicx}  
\usepackage{subeqnar}  
\usepackage{multicol}  
\usepackage{physprbb}  
\makeindex             

\begin{document}
\title*{[O\,{\sc i}] 6300 \AA\  Emission from Disks Around Herbig Ae/Be Stars}
\toctitle{[O\,{\sc i}] 6300 \AA\  Emission from Disks Around Herbig Ae/Be Stars}
\titlerunning{[O\,{\sc i}] 6300 \AA\  Emission from Disks Around Herbig Ae/Be Stars}
%
\author{M.E. van den Ancker\inst{1} \and B Acke\inst{2} \and C.P. Dullemond\inst{3}}
\authorrunning{M.E. van den Ancker et al.}
%
%
\institute{
European Southern Observatory, Karl-Schwarzschild-Str. 2, D--85748 
 Garching bei M\"unchen, Germany
\and 
Instituut voor Sterrenkunde, Katholieke Universiteit Leuven, 
 Celestijnenlaan 200B, 3001 Leuven, Belgium
\and 
 Max-Planck-Institut f\"ur Astrophysik, Postfach 1317, D--85748  Garching bei M\"unchen, Germany}

\maketitle              

\begin{abstract}
We present high spectral-resolution optical spectra of 49 Herbig Ae/Be stars in a search for the [O\,{\sc i}] 6300.2 \AA\  line.  
The vast majority of the stars in our sample show narrow (FWHM $<$ 100 km~s$^{-1}$) emission lines, centered 
on the stellar radial velocity.   Some stars in our sample show double-peaked lines profiles, with peak-to-peak 
separations of $\sim$ 10 km~s$^{-1}$.  The presence and strength of the [O\,{\sc i}] line emission appears to be well correlated 
with the far-infrared energy distribution of each source: stars with a strong excess at 60 $\mu$m have in general 
stronger [O\,{\sc i}] emission than stars with weaker 60~$\mu$m excesses. We interpret  the [O\,{\sc i}] 6300.2 \AA\  line profiles 
as arising in the surface layers of the protoplanetary disks surrounding Herbig Ae/Be stars.   
\end{abstract}

\section{Observations}
High-resolution optical spectra of 49 Herbig Ae/Be stars were obtained with the Coud\'e Echelle Spectrograph 
($R$ = 130,000) on the ESO 3.6 m telescope, the Echelle spectrograph on the Mayall 4m telescope at KPNO ($R$ = 30,000) 
and the Utrecht Echelle Spectrograph on the William Herschel Telescope ($R$ = 45,000).  The spectra were examined 
for the presence of the [O\,{\sc i}] 6300.2 \AA\  line -- the strongest forbidden emission line at optical wavelengths.  
We detect [O\,{\sc i}] 6300.2 \AA\  in emission in 39 out of the 49 sources in our sample.  The vast majority of these 
sources show narrow (FWHM $<$ 100 km~s$^{-1}$) profiles, centered on the stellar radial velocity or slightly blue-shifted 
(centroid position up to $-$70 km~s$^{-1}$).  In only three sources in which [O\,{\sc i}] emission was detected, the feature is 
much broader ($\sim$ 400 km~s$^{-1}$) and strongly blue-shifted ($\sim$ $-$200 km~s$^{-1}$) with respect to the 
stellar radial velocity.    
\begin{figure}[t]
\begin{center}
\includegraphics[width=.45\textwidth,angle=90]{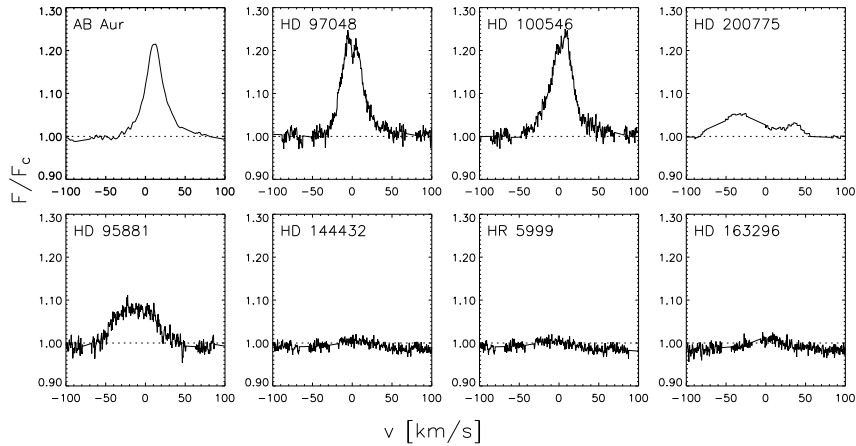}
\end{center}
\caption[]{Examples of [O\,{\sc i}] spectra of Herbig Ae/Be stars. Areas of the spectrum strongly affected by telluric 
        absorption or [O\,{\sc i}] airglow have been removed from these spectra.  The horizontal and vertical 
        lines indicate the continuum level and stellar radial velocity.}
\label{eps1}
\end{figure}

\section{Analysis of Emission Lines}
A more careful inspection of the centroid velocities of the detected [O\,{\sc i}] emission lines shows that the vast 
majority of Herbig Ae/Be stars have centroid velocities that are close to the stellar radial velocity.  Nevertheless, 
there appears to be a small minority of Herbig stars that show slightly (up to $-$70 km~s$^{-1}$) blue-shifted centroid 
velocities as well (Fig. 2).  

In general, the lines -- all of which are resolved at our spectral resolution -- are narrow (FWHM $<$ 100 km~s$^{-1}$), 
without any obvious dependence of line width on line strength (Fig. 3).  A noticeable exception to this 
general rule are the three embedded sources PV Cep, V645 Cyg and Z CMa, which show strong, much broader 
(FWHM $\sim$ 400 km~s$^{-1}$) and strongly blue-shifted [O\,{\sc i}] emission lines. 

Our high spectral resolution and high S/N spectra allow us to resolve the line profiles exhibited by the 
narrow emission components.  Although the line profiles in most stars are single-peaked and symmetric, there 
are clear examples in our sample of stars showing double-peaked line profiles with peak-to-peak separations 
of $\sim$ 10 km~s$^{-1}$, with the wings of the spectral line extending to $\sim$ 40 km~s$^{-1}$.  If we assume that these profiles 
are due to material that is in Keplerian rotation around the star, these lines would be formed at radii of 
1--60 AU from the central star.  These are exactly the types of profiles and radii that would correspond to 
a location in the protoplanetary disks surrounding Herbig Ae/Be stars.

\begin{figure}[t]
\begin{center}
\includegraphics[width=.60\textwidth,angle=0]{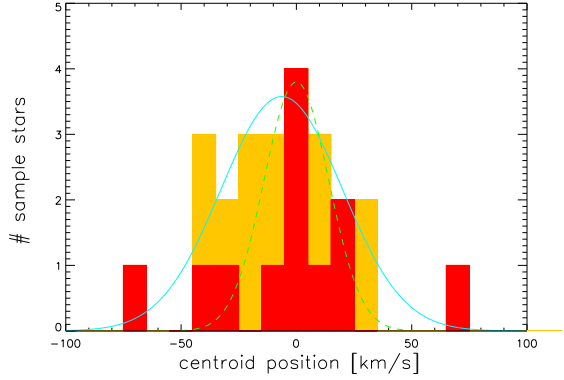}
\end{center}
\caption[]{The histogram of the centroid positions for the pure emission profiles in our sample. The bin size 
        is 10 km~s$^{-1}$.  The dashed line represents the distribution expected if all sources were centered on 
        the stellar radial velocity; the solid line is a Gaussian fit to this histogram.  Yellow parts 
        of the bars indicate group I sources, red group II.}
\label{eps2}
\end{figure}

\begin{figure}[t]
\begin{center}
\includegraphics[width=.60\textwidth,angle=0]{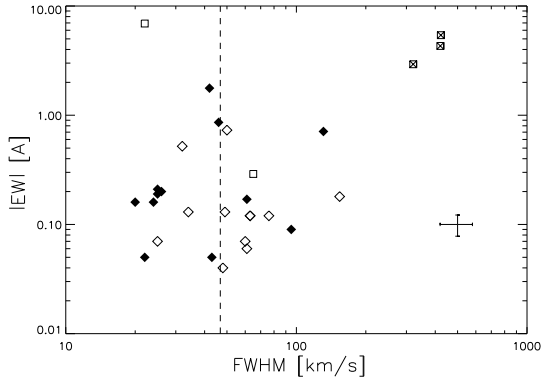}
\end{center}
\caption[]{The Equivalent Width (EW) of the [O\,{\sc i}] emission lines versus the FWHM.  Filled diamonds represent 
        group I, diamonds group II and squares group III sources.  The x-marked squares represent PV Cep, 
        V645 Cyg and Z CMa.  The dashed line indicates the mean FWHM (47 km~s$^{-1}$).}
\label{eps3}
\end{figure}

\section{Correlations with Other Properties}
One of the more striking features of the [O\,{\sc i}] emission exhibited by the stars in our sample is the large 
range in  [O\,{\sc i}] emission line strength.  Although stars with higher luminosities display on average also 
stronger [O\,{\sc i}] emission, for stars with [O\,{\sc i}] 6300.2 \AA\  line luminosities below a few times 10$^{-6}$~L$_\odot$ there 
are roughly equal numbers of stars in which we detect and do not detect significant [O\,{\sc i}] emission, 
regardless of stellar luminosity.  This behaviour is reminiscent of what was recently found for the 
infrared emission bands due to Polycyclic Aromatric Hydrocarbons (PAHs) (Acke \& van den Ancker 2004).  
Remarkably, the correlation between PAH and [O\,{\sc i}] luminosity appears fairly tight (Fig. 5).

\begin{figure}[t]
\begin{center}
\includegraphics[width=.60\textwidth,angle=0]{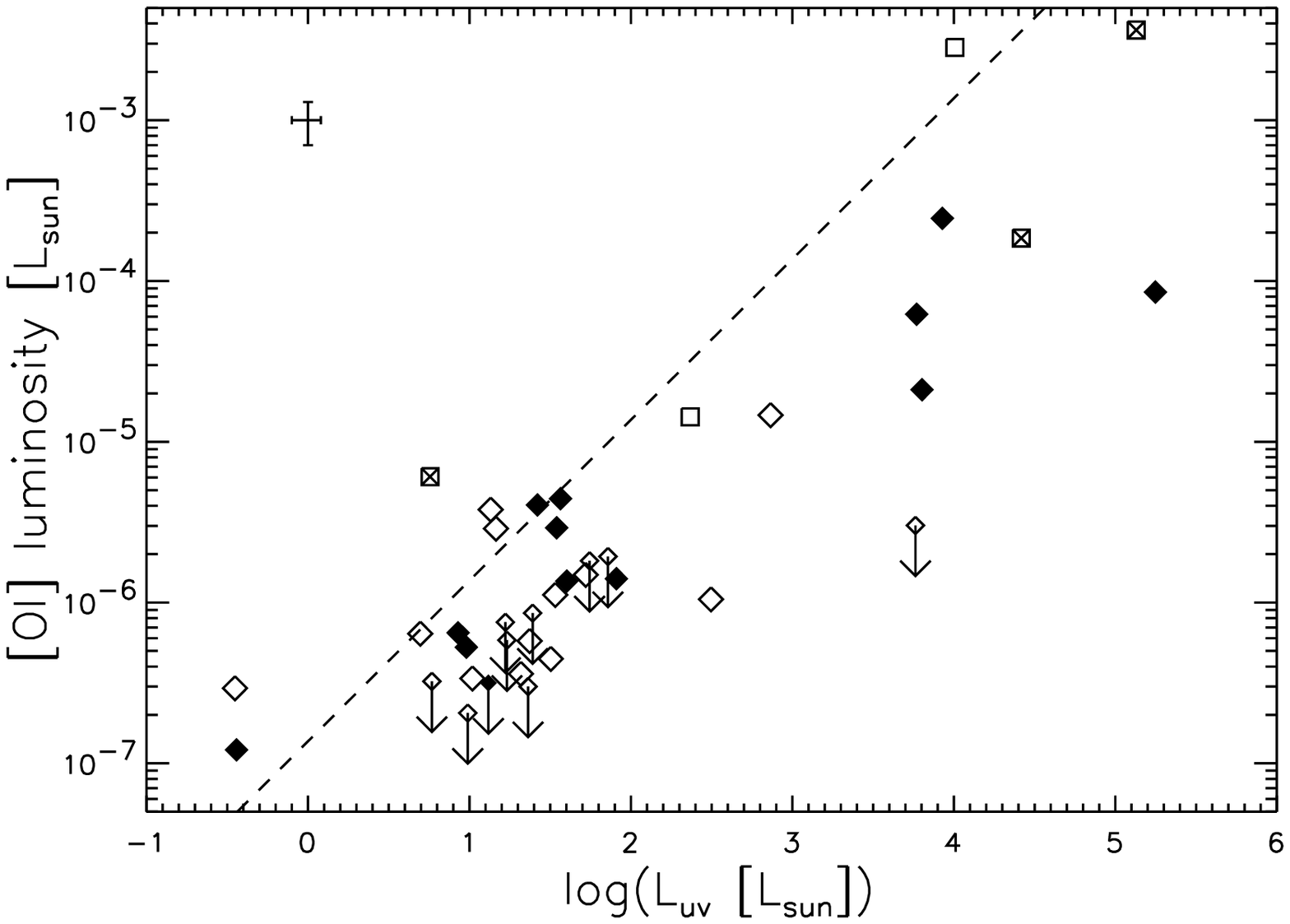}
\end{center}
\caption[]{The [O\,{\sc i}] 6300 \AA\  luminosity versus the stellar UV luminosity.  The plotting symbols are 
        as in Fig. 3. A typical error bar is indicated in the upper left corner.  The dashed line 
        represents L([O\,{\sc i}]) = 1.4 $\times$ 10$^{-7}$~L$_{\rm UV}$.}
\label{eps3}
\end{figure}
\begin{figure}[t]
\begin{center}
\includegraphics[width=.60\textwidth,angle=0]{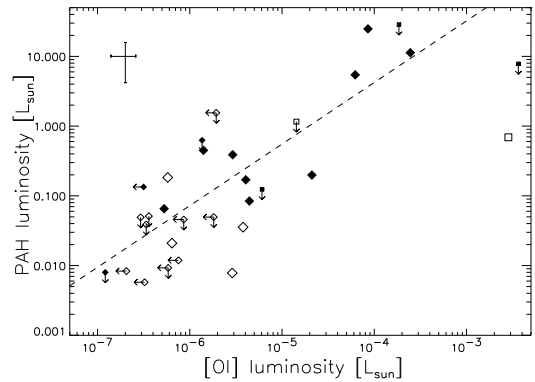}
\end{center}
\caption[]{The PAH luminosity versus the [O\,{\sc i}] 6300 \AA\  luminosity. The dashed line represents 
        the mean PAH-over-[O\,{\sc i}] luminosity ratio (L(PAH)/L[O\,{\sc i}] $\approx$ 10$^5$).}
\label{eps5}
\end{figure}

\section{Discussion and Conclusions}
As shown above, the observed [O\,{\sc i}] 6300.2 \AA\  emission is believed to originate from the circumstellar disk. 
Typical disk temperatures are much too low to explain the strength of the observed lines through thermal 
emission.  However, the chemical reaction OH + H $\rightarrow$ O + H$_2$ leaves a significant fraction of the produced 
Oxygen atoms in the $^1D_2$ state; the upper level of the 6300.2 \AA\  line (e.g. St\"orzer \& Hollenbach 2000).  
We suggest here that the bulk of the observed [O\,{\sc i}] emission is non-thermal and produced 
by the photo-dissociation of OH.  Since this latter process critically depends on the ability of stellar 
UV photons to reach the OH molecule, we expect disk geometry to have a significant effect on the strength 
of the [O\,{\sc i}] 6300.2 \AA\  line in Herbig Ae/Be stars.

The spectral energy distribution (SED) of Herbig stars can be roughly classified in two groups: group I 
are sources with a strong mid-IR (20--100~$\mu$m) excess, group II have more modest mid-IR emission 
(Meeus et al. 2001). Dullemond (2002) interprets this classification in terms 
of disk geometry: group I sources have flared disks, group II self-shadowed disks (Fig. 6). The 
classification of the sources can be expressed in a diagram such as shown in Fig. 7. The strong mid-IR 
group I sources are redder than the group II sources and appear in the lower right part of this diagram.  
Group II sources are clearly more moderate [O\,{\sc i}] emitters than group I sources. The flaring of the disk 
in group I sources makes that the surface layers of the disk directly see the stellar UV photons 
capable of photo-dissociating OH. In group II sources, the outer parts of the disk lie in the shadow 
of the puffed-up inner rim and hence receive almost no direct stellar radiation. The observed strong 
difference in [O\,{\sc i}] 6300.2 \AA\  emission properties is therefore consistent with our previous 
interpretation of the [O\,{\sc i}] emission as being due to the photo-dissociation of OH.
 
\begin{figure}[t]
\begin{center}
\includegraphics[width=.25\textwidth,angle=270]{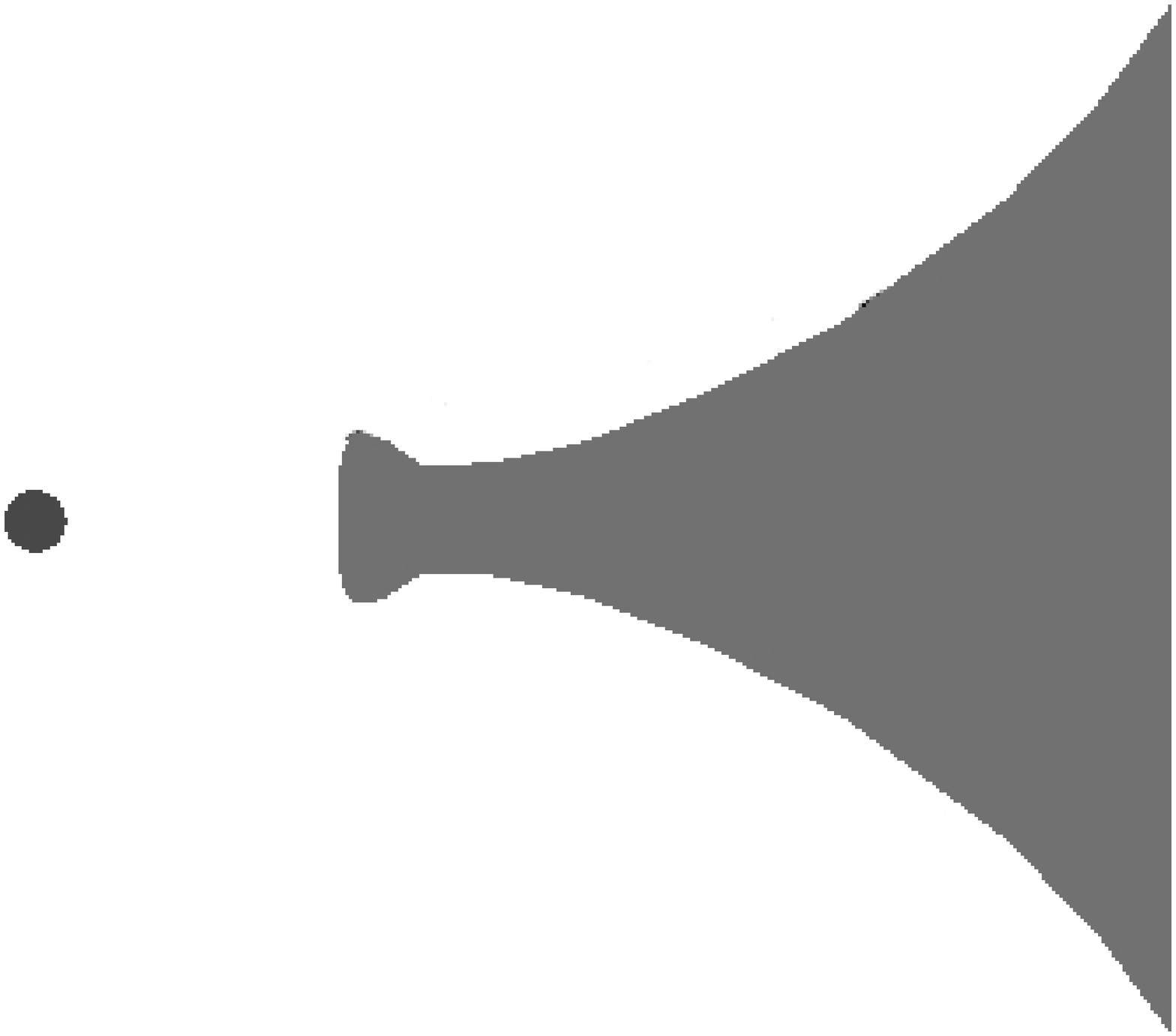}\hspace*{2.0cm}
\includegraphics[width=.25\textwidth,angle=270]{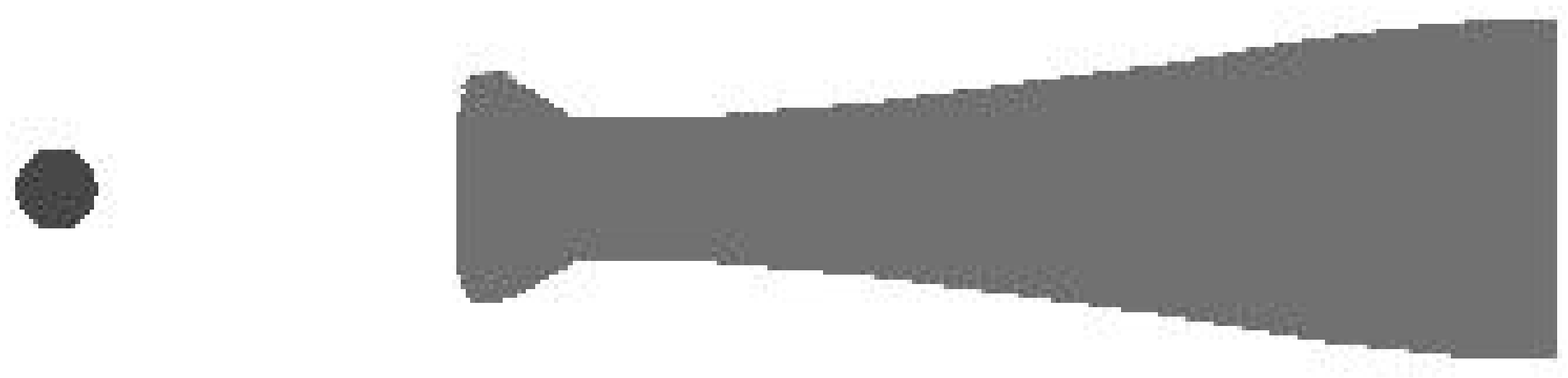}
\end{center}
\begin{center}
\includegraphics[width=.25\textwidth,angle=270]{hd100453_sed.ps}\hspace*{0.5cm}
\includegraphics[width=.25\textwidth,angle=270]{hd31648_sed.ps}
\end{center}
\caption[]{Above: pictograms of the two different types of disk models: flared (left) and self-shadowed. 
        Below: the SEDs of two sample stars are shown in order to illustrate the difference between 
        group I and group II. Stars with flared disks have group I SEDs, self-shadowed-disk systems 
        are group II objects. The puffed-up inner rim of the disk causes the near-IR bump in the SEDs, 
        while the flaring outer parts lead to the strong mid-IR excess in group I sources.}
\label{eps6}
\end{figure}
\begin{figure}
\begin{center}
\includegraphics[width=.60\textwidth,angle=0]{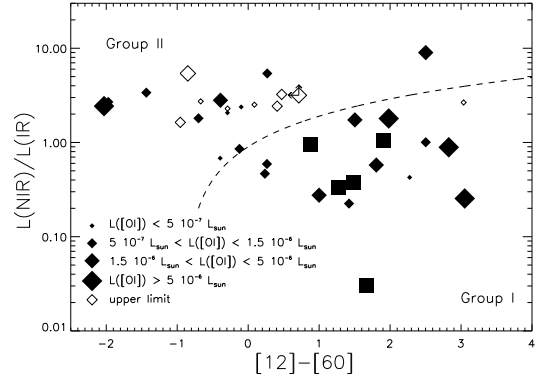}
\end{center}
\caption[]{The ratio of the near-to-mid-IR luminosity $L_{\rm NIR}/L_{\rm IR}$ is plotted versus the 
        IRAS [12]$-$[60] color (after van Boekel et al. 2003). The plotting symbols are 
        proportional to the strength of the [O\,{\sc i}] luminosity. Filled symbols indicate detected 
        features, open symbols refer to upper limits. Squares represent highly embedded objects.}
\label{eps7}
\end{figure}

%

\end{document}